\documentclass[twocolumn,english,showpacs,prl,floatfix]{revtex4}
\usepackage[]{fontenc}
\usepackage[latin1]{inputenc}
\usepackage{amsmath}
\usepackage{graphicx}
\usepackage{amssymb}

\makeatletter

\usepackage{bm}

\usepackage{babel}
\makeatother

\begin{document}

\title{On a Proper Definition of Spin Current}
\author{Junren Shi}
\thanks{These authors contributed equally to this work.}

\author{Ping Zhang}
\thanks{These authors contributed equally to this work.}

\author{Di Xiao}

\author{Qian Niu}

\affiliation{Department of Physics, The University of Texas at Austin,
  Austin, TX 78712}

\begin{abstract}
  The conventional definition of spin current is incomplete and
  unphysical in describing spin transport in systems with spin-orbit
  coupling.  A proper and measurable spin current is established in
  this study, which fits well into the standard framework of
  near-equilibrium transport theory and has the desirable property to
  vanish in insulators with localized orbitals.  Experimental
implications of our theory are discussed.
\end{abstract}


\maketitle
A central theme of spintronics research is on how to generate
and manipulate spin current as well as to exploit its various
effects~\cite{Wolf,Das}. In the ideal situation where spin (or its
projection along a direction) is conserved, spin current is simply
defined as the difference between the currents of electrons in the two
spin states. This concept has served well in early study of
spin-dependent transport effects in metals. The ubiquitous presence of
spin-orbit coupling inevitably makes the spin non-conserved, but this
inconvenience is usually put off by focusing one's attention within
the so-called spin relaxation time. In recent years, it has been found
that one can make very good use of spin-orbit coupling, realizing
electric control of spin generation and transport
~\cite{Dya,Hirsch,Zhang,Muk1,Sinova,Kato,Wund}. The question of how to
define the spin current properly in the general situation therefore
becomes urgent.

\begin{figure}
  \includegraphics[width=0.8\columnwidth]{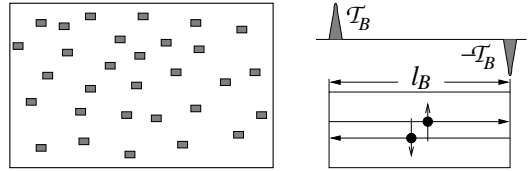}
  \caption{\label{spinbox} An example demonstrating the irrelevance of
    the conventional spin current to spin transport. Left: A
    macroscopic system consisting of a uniform distribution of
    microscopic boxes (gray rectangles) in which an electron is
    confined.  Right: The inside structure of the box, where the walls
    flips the spin of electron upon collision.  When the system is
    driven to such a state schematized at the right, the macroscopic
    average of the conventional spin current is nonzero.  However,
    because all electrons are localized, no spin transport or boundary
    spin accumulation can occur.  $\mathcal{T}_B $ denotes the total
    boundary spin torque due to the spin-flip scattering.  It is easy
    to show the corresponding spin torque dipole
    $P_\tau=-\mathcal{T}_B l_B$ exactly cancels the conventional spin
    current.  As a result, the spin current $\bm{\mathcal{J}}_s$
    defined in Eq.~(\ref{eq:conserved spin current}) is zero.}
\end{figure}

In most previous studies of bulk spin transport, it has been
conventional to define the spin current simply as the expectation
value of the product of spin and velocity observables.  Unfortunately,
no viable measurement is known to be possible for this spin current.
The recent spin-accumulation experiments~\cite{Kato,Wund} do not
directly determine it, and there is no deterministic relation between
this spin current and the boundary spin accumulation, as demonstrated
in Fig.~\ref{spinbox}. 

In fact, the conventional definition of spin current suffers three
critical flaws that prevent it from being relevant to spin transport.
First, this spin current is not conserved. This issue alone has
motivated a number of alternative definitions
recently~\cite{Muk2,Zhang2005,Jin2005}. Second, this spin current can
even be finite in insulators with localized eigenstates only, so it
cannot really describe transport~\cite{Rashba2003}.  Finally, there
does not exist a mechanical or thermodynamic force in conjugation with
this current, so it cannot be fitted into the standard
near-equilibrium transport theory. One consequence is that one cannot
establish an Onsager relation linking the spin current with other
transport phenomena.

In this Letter, we try to establish a proper definition of spin
current free from all the above difficulties, which is found to be
possible for systems where spin generation in the bulk is absent due
to symmetry reasons.  Our new spin current is given by the time
derivative of the spin displacement (product of spin and position
observables), which differs from the conventional definition by a
torque dipole term.  The torque dipole term is first found in a
semiclassical theory~\cite{Culcer}, whose impact on spin transport has
been further analyzed to assess the importance of the intrinsic spin
Hall effect \cite{Zhang2005}.  In this work, we provide a quantum
mechanical description of this term within the linear response theory
for transport.  Apart from showing its consequence in making the spin
current conserved, we also reveal two additional properties: The new
spin current vanishes identically in insulators with localized
orbitals, and is in conjugation with a force given by the gradient of
the Zeeman field or spin-dependent chemical potential.  Together with
conservation, these properties are crucial to establish the new spin
current as the proper description for spin transport, and they also
provide a firm foundation for various methods for its measurement.
\\

Based on general quantum mechanical principle, one can derive a
continuity equation relating the spin, current and torque densities as
follows,
\begin{eqnarray} 
  \frac{\partial S_{z}}{\partial
    t}+\nabla\cdot\mathbf{J}_{s} & = &
  \mathcal{T}_{z}\,.\label{eq:continuity}
\end{eqnarray} 
The spin density for a particle in a (spinor) state $\psi(\mathbf{r})$
is defined by
$S_{z}(\mathbf{r})=\psi^{\dagger}(\mathbf{r})\hat{s}_{z}\psi(\mathbf{r})$,
where $\hat{s}_{z}$ is the spin operator for a particular component
($z$ here, to be specific). The spin current density here is given by
the conventional definition
$\mathbf{J}_{s}(\mathbf{r})=\mathrm{Re}\psi^{\dagger}(\mathbf{r})\frac{1}{2}
\{\hat{\mathbf{v}},\,\hat{s}_{z}\}\psi(\mathbf{r})$, where
$\hat{\mathbf{v}}$ is the velocity operator, and $\{,\}$ denotes the
anticommutator. The right hand side of the continuity equation is the
torque density defined by
$\mathcal{T}_{z}(\mathbf{r})=\mathrm{Re}\psi^{\dagger}(\mathbf{r})\hat{\tau}
\psi(\mathbf{r})$, where
$\hat{\tau}\equiv\mathrm{d}\hat{s}_{z}/\mathrm{d}t\equiv(1/i\hbar)
[\hat{s}_{z},\,\hat{H}]$, and $\hat{H}$ is the Hamiltonian of the
system. These definitions can be easily restated in a many-body
language by regarding the wave functions as field operators and by
taking expectation value in the quantum state of the system. The
presence of the torque density $\mathcal{T}_{z}$ reflects the fact
that spin is not conserved microscopically in systems with spin-orbit
coupling.

It often happens, due to symmetry reasons, that the average torque
vanishes for the bulk of the system, i.e., $(1/V)\int
dV\mathcal{T}_{z}(\mathbf{r})=0$.  This is true to first order in the
external electric field for any samples with inversion symmetry. Also,
one is often interested in a particular component of the spin, and the
corresponding torque component can vanish in the bulk on average even
for samples without the inversion symmetry. This is certainly true for
the many models used for the study of spin Hall
effect~\cite{Sinova,Sch1,Muk1}, and for the experimental systems used
to detect the effect so far ~\cite{Kato,Wund}. For such systems, where
the average spin torque density vanishes in the bulk, we can write the
torque density as a divergence of a torque dipole density,
\begin{equation}
  \mathcal{T}_{z}(\mathbf{r})=-\nabla\cdot\mathbf{P}_{\tau}(\mathbf{r})
  \,.\label{eq:spin torque density}
\end{equation} 
Moving it to the left hand side of (\ref{eq:continuity}), we have
\begin{eqnarray}
  \frac{\partial S_{z}}{\partial t}+\nabla\cdot\left(\mathbf{J}_{s}+
    \mathbf{P}_{\tau}\right) & = & 0\,,\label{eq:continuity1}
\end{eqnarray}
which is in the form of the standard sourceless continuity equation.
This shows that the spin in conserved on average in such systems, and the
corresponding transport current is:
\begin{eqnarray} 
  \bm{\mathcal{J}}_{s} & = & \mathbf{J}_{s}+\mathbf{P}_{\tau}.
  \label{eq:conserved spin current}
\end{eqnarray}

We note that there is still an arbitrariness in defining the effective
spin current because Eq.~(\ref{eq:spin torque density}) does not
uniquely determine the torque dipole density $\mathbf{P}_{\tau}$ from
the corresponding torque density $\mathcal{T}_{z}$. We can eliminate
this ambiguity by imposing the physical constraint that the torque
dipole density is a material property that should vanish outside the
sample. This implies in particular that $\int
dV\,\mathbf{P}_{\tau}=-\int
dV\,\mathbf{r}\nabla\cdot\mathbf{P}_{\tau}=\int
dV\,\mathbf{r}\mathcal{T}_{z}(\mathbf{r})$.  It then follows that,
upon bulk average, the effective spin current density can be written
in the form of
$\boldsymbol{\mathcal{J}}_{s}=\mathrm{Re}\psi^{*}(\mathbf{r})
\hat{\boldsymbol{\mathcal{J}}}_{s}\psi(\mathbf{r})$, where
\begin{eqnarray} 
  \hat{\boldsymbol{\mathcal{J}}}_{s} & = & \frac{\mathrm{d}
    (\hat{\mathbf{r}}\hat{s}_{z})}{\mathrm{d}t}\,
  \label{eq:conserved spin current operator}
\end{eqnarray} 
is the effective spin current operator. Compared to the conventional
spin current operator, it has an extra term
$\hat{\mathbf{r}}(\mathrm{d}\hat{s}_{z}/\mathrm{d}t)$, which accounts
the contribution from the spin torque.

Because the new spin current is given as a time derivative, 
it must vanish in an eigen energy state in which the spin 
displacement operator is well defined, which is the case if the state 
is localized.  Elementary perturbation theory shows that the spin current  
vanishes in such a system even in the presence of a weak electric field.
Indeed, for spatially localized eigenstates, we can evaluate the spin
transport coefficient as,
\begin{eqnarray}
  \sigma^{s} & = & -e\hbar\sum_{\ell\ne\ell'}f_{\ell}
  \frac{\mathrm{Im}\langle\ell|\mathrm{d}(\hat{\mathbf{r}}
    \hat{s}_{z})/\mathrm{d}t|\ell'\rangle\langle\ell'|
    \hat{\mathbf{v}}|\ell\rangle}
  {(\epsilon_{\ell}-\epsilon_{\ell'})^{2}}
  \label{eq:sigmas for insulator}\\
  & = &
  -e\hbar\sum_{\ell}f_{\ell}\langle\ell|[\hat{\mathbf{r}}
  \hat{s}_{z},\,\hat{\mathbf{r}}]|\ell\rangle=0
  \nonumber
\end{eqnarray} 
where $f_{\ell}$ is the equilibrium occupation number in the $\ell$-th
state. Here, we have used
$\langle\ell|\mathrm{d}(\hat{\mathbf{r}}\hat{s}_{z})/
\mathrm{d}t|\ell'\rangle=(-i/\hbar)(\epsilon_{\ell'}-\epsilon_{\ell})
\langle\ell|\hat{\mathbf{r}}\hat{s}_{z}|\ell'\rangle$ and
$\langle\ell'|\hat{\mathbf{v}}|\ell\rangle=(-i/\hbar)
(\epsilon_{\ell}-\epsilon_{\ell'})\langle\ell'|\hat{\mathbf{r}}|\ell\rangle$.
The involved matrix elements are all well defined between spatially
localized eigenstates.

Defined as a time derivative of the spin displacement operator
$\hat{\mathbf{r}}\hat{s}_z$, the new spin current has a natural
conjugate force $\mathbf{F}_{s}$, the gradient of the Zeeman field or
of a spin dependent chemical potential, which can be modeled as an
external perturbation
$V=-\mathbf{F}_{s}\cdot(\hat{\mathbf{r}}\hat{s}_{z})$~\cite{Zutic2002}.
The energy dissipation rate for the spin transport can be
written as $\mathrm{d}Q/\mathrm{d}t =
\bm{\mathcal{J}}_s\cdot\mathbf{F}_s $. It immediately suggests a
thermodynamic way to determine the spin current by simultaneously
measuring the Zeeman field gradient (spin force) and the heat
generation.

Moreover, Onsager relations can now be established. For example, in
the presence of both electric and spin forces, the linear response of
spin and charge currents may be written in the following manner,
\begin{equation}
  \left(\begin{array}{c}
      \bm{\mathcal{J}}_{s}\\
      \mathbf{J}_{c}\end{array}\right)=\left(\begin{array}{cc}
      \sigma^{ss} & \sigma^{sc}\\
      \sigma^{cs} & \sigma^{cc}\end{array}\right)\left(\begin{array}{c}
      \mathbf{F}_{s}\\
      \mathbf{E}\end{array}\right),\label{eq:Transport Matrix}
\end{equation}
where $\bm{\mathcal{J}}_{s}$ is spin current and $\mathbf{J}_{c}$
denotes charge current. $\sigma^{ss}$ and $\sigma^{cc}$ are the
spin-spin and charge-charge conductivity tensors respectively. The off
diagonal block $\sigma^{sc}$ denotes spin current response to an
electric field (spin Hall effect), and $\sigma^{cs}$ denotes charge
current response to a spin force (inverse spin-Hall
effect)~\cite{PZhang}.  The Onsager reciprocity dictates a general
relation between the off-diagonal blocks (assuming time reversal
symmetry):
\begin{equation}
  \sigma_{\alpha\beta}^{sc}=-\sigma_{\beta\alpha}^{cs}\label{eq:Onsage Relation}
\end{equation}
where the extra minus sign originates from the odd time-reversal
parity of the spin displacement operator
$\mathbf{r}s_{z}$~\cite{Casimir}.  We note that the Onsager relation
Eq.~(\ref{eq:Onsage Relation}) can only be established when the
currents are defined in terms of the time derivative of displacement
operators conjugate to the forces.  In the case of spin force driving,
we have the spin displacement operator $s_{z}\mathbf{r}$, whose time
derivative corresponds to our new spin current Eq.~(\ref{eq:conserved
  spin current operator}), not the conventional spin current. This
Onsager relation
can also be directly verified using the linear response 
theory.

The existence of the Onsager relations makes the electric
measurement of the spin current viable.  Previous theoretical proposal
had suggested that the spin current can be determined by measuring the
transverse voltage generated by a spin current passing through a spin
Hall device~\cite{Hirsch,Zhang,Han}. However, to do that, the spin
Hall coefficient $\sigma^{sc}_{xy}$ of the measuring device must be
known in prior because the spin current is determined from $
\mathcal{J}_{sx} = \sigma^{sc}_{xy} E_y $.  With the Onsager relation,
$\sigma^{sc}_{xy}$ can be derived from the corresponding inverse spin
Hall coefficient, and the latter can be determined by a measurement of
the charge current and the Zeeman field gradient. 
\\

After these general considerations, we now show how to evaluate 
the spin Hall conductivity based on the new definition of spin current.  
The torque dipole density can be determined unambiguously as a bulk
property within the theoretical framework of linear response. Consider
the torque response to an electric field at finite wave vector
$\textbf{q}$,
$\mathcal{T}_{z}(\mathbf{q})=\boldsymbol{\chi}(\mathbf{q})
\cdot\mathbf{E}(\mathbf{q})$.  Based on Eq.~(\ref{eq:spin torque
  density}) which implies
$\mathcal{T}_{z}(\mathbf{q})=-i\mathbf{q}\cdot\mathbf{P}_{\tau}(\mathbf{q})$,
we can uniquely determine the dc response (i.e.,
$\mathbf{q}\rightarrow0$) of the spin torque dipole:
\begin{eqnarray}
  \mathbf{P}_{\tau} & = & \mathrm{Re}\{i\nabla_{\mathbf{q}}
  [\boldsymbol{\chi}(\mathbf{q})\cdot\mathbf{E}]\}_{\mathbf{q}=0}\,.
  \label{eq:spin torque dipole}
\end{eqnarray} 
Here we have utilized the condition $\boldsymbol{\chi}(0)=0$, i.e.,
there is no bulk spin generation by the electric field. Combining
Eq.~(\ref{eq:conserved spin current}) and (\ref{eq:spin torque
  dipole}), we can then determine the electric-spin transport
coefficients for the new definition of spin current:
\begin{eqnarray}
  \sigma_{\mu\nu}^{s} & = &
  \sigma_{\mu\nu}^{s0}+\sigma_{\mu\nu}^{\tau}\,,\label{eq:spin
    transport coefficient}
\end{eqnarray} 
where $\sigma_{\mu\nu}^{s0}$ is the conventional spin-transport
coefficient that is the focus of most of previous studies, and
\begin{equation}
  \sigma_{\mu\nu}^{\tau}=\mathrm{Re}[i\partial_{q_{\mu}}
  \chi_{\nu}(\mathbf{q})]_{\mathbf{q}=0}\,,
  \label{eq:sigmatau}
\end{equation}
is the contribution from the spin torque dipole.  Standard Green
function or many-body techniques can be used to evaluate this linear
response for systems with arbitrary disorder and interactions between
the carriers.

The spin-Hall coefficients for a few semiconductor models including
effects of disorder are now considered by Sugimoto et al. based on our
new definition of spin current~\cite{Sugimoto2005}, who found results
dramatically different from the conventional spin Hall
conductivities.  For example, it is found that the spin
Hall conductivity depends explicitly on the scattering potentials for
the two dimensional Rashba models with k-linear or k-cubic spin-orbit
coupling.  For the k-cubic model, the conventional spin Hall
conductivity is robust against disorder, but this is not so if the new
spin current definition is adopted.  It then implies that at smooth
boundaries, spin accumulation in such systems are of extrinsic nature.


%
%

The conservation of our new spin current allows one to consider spin
transport in the bulk without the need of laboring explicitly a spin
torque (dipole density) which may be generated by the electric field.
One can think of spin transport for systems with strong spin-orbit
coupling in the ``usual'' sense established for systems with weak
spin-orbit coupling.  For example, it has been customary to link spin
density and spin current through the following phenomenological
equation of spin continuity,
\begin{eqnarray}
  \frac{\partial S_{z}}{\partial t}+\nabla\cdot \bm{\mathcal{J}}_{s} 
  & = & - {S_{z} \over \tau_s}\,,\label{eq:continuity2}
\end{eqnarray}
where $\tau_s$ is the spin relaxation time, and the spin current has
the form $\bm{\mathcal{J}}_{s}=\sigma E - D_s \nabla S_z$. This makes
sense only if our new spin current is used in the calculation of spin
Hall conductivity $\sigma$, otherwise an extra term of field-generated
spin torque must be added~\cite{Note1}.

Equation (\ref{eq:continuity2}) can serve as the basis to determine
the spin accumulation at a sample boundary, which is of much current
interest. Consider a system having a smooth boundary produced by a
slowly varying confining potential.  We assume that the length scale
of variation is much larger than the mean free path, so that the above
continuity equation may be applied locally.  By integrating from the
interior to the outside of the sample boundary, we obtain a spin
accumulation per area with
$\bar{S}_{z}=\mathcal{J}_{s}^{\textrm{bulk}}\tau_{s}$~\cite{Culcer},
where $\tau_s$ is the spin relaxation time.  We emphasize that the
transport spin current responsible for the boundary spin accumulation
should be $\bm{\mathcal{J}}_s$ instead of the conventional spin
current $\bm J_s$.  

For sharp boundaries, the spin continuity equation alone cannot yield
a unique relationship between the spin current from the bulk and spin
accumulation at the boundary.  For a perfectly reflecting sharp wall
in the case of strong spin-orbit coupling, the boundary spin
accumulation seems to be determined by the conventional spin current
from the bulk \cite{Nomura2005}.  However, such a relationship is
altered for other boundary conditions~\cite{Tse2005}.  This calls for
well controlled experiments to explicitly eliminate the influence from
the boundary condition. On the other hand, for the generic class of
smooth boundaries discussed above, there is a unique relationship
between spin accumulation and the spin current, provided one uses our
new definition.

The real advantage of our new definition of spin current lies in the
fact that it provides a satisfactory description of spin transport in
the bulk.  With our new spin current, one can now use the spin
continuity equation (12) to discuss spin accumulation in the bulk,
e.g., by generating a non-uniform electric field or spatially
modulating the spin Hall conductivity.  Our new spin current vanishes
in Anderson insulators either in equilibrium or in a weak electric
field, which enables us to predict zero spin accumulation in such
systems.  More importantly, it posses a conjugate force (spin force),
so that spin transport can be fitted into the standard formalism of
near equilibrium transport.  The conventional spin current does not
have a conjugate force, so it makes no sense even to talk about energy
dissipation from that current.  The existence of a conjugate force is
crucial for the establishment of Onsager relations between spin
transport and other transport phenomena, and its measurement will be
important to thermodynamic and electric determination of the spin
current.

We gratefully acknowledge discussions with S. Zhang, Z. Yang, D.
Culcer, J. Sinova, S.-Q.  Shen and A.H. MacDonald.  QN and DX were
supported by DOE (DE-FG03-02ER45958) and the Welch Foundation, and JRS
and PZ were supported by NSF (DMR-0306239 and DMR 0404252).

\end{document}